\documentclass[sigplan,screen,nonacm]{acmart}

\usepackage{tabularx}
\AtBeginDocument{%
  \providecommand\BibTeX{{%
    \normalfont B\kern-0.5em{\scshape i\kern-0.25em b}\kern-0.8em\TeX}}}

\acmConference[BiblioDAP’21]{The 1st Workshop on Bibliographic Data Analysis and Processing}{August 14--18, 2021}

\begin{document}

\title{Multimodal Approach for Metadata Extraction from German Scientific Publications}


\author{Azeddine Bouabdallah \qquad Jorge Gavilan \qquad Jennifer Gerbl \qquad Prayuth Patumcharoenpol}
\authornote{All authors contributed equally to this research.}
\email{{bazeddine, jorge, jgerbl, prayuthp}@uni-koblenz.de}
\affiliation{%
  \institution{\textit{Institute for Web Science and Technologies (WeST)\\
  University of Koblenz-Landau}}
  \city{Koblenz}
  \country{Germany}
}




\renewcommand{\shortauthors}{Azeddine Bouabdallah, Jorge Gavilan, Jennifer Gerbl and Prayuth Patumcharoenpol}

\begin{abstract}
Nowadays, metadata information is often given by the authors themselves upon submission. However, a significant part of already existing research papers have missing or incomplete metadata information. German scientific papers come in a large variety of layouts which makes the extraction of metadata a non-trivial task that requires a precise way to classify the metadata extracted from the documents. In this paper, we propose a multimodal deep learning approach for metadata extraction from scientific papers in the German language. We consider multiple types of input data by combining natural language processing and image vision processing. This model aims to increase the overall accuracy of metadata extraction compared to other state-of-the-art approaches. It enables the utilization of both spatial and contextual features in order to achieve a more reliable extraction. Our model for this approach was trained on a dataset consisting of around 8800 documents and is able to obtain an overall F1-score of 0.923.
\end{abstract}




\maketitle

\section{Introduction}
In recent decades, metadata extraction has become an important task in many fields. This importance arises from the increase in digital content usage, as more people are now used to obtaining information through the internet. Consequently, it increased the need of providing easy access and retrieval of information for users. Ensuring this requires the availability of contents' metadata to help retrieval systems in finding relevant content. Digital scholar libraries are one of the major parties needing to keep this easy access constantly. Recent statistics estimated that there are around 1.8 million scientific publications each year \cite{stmreport}, most of which are of a digital format. This high number of publications each year and past publications introduce a new challenge in acquiring their metadata. 

Currently, digital libraries are relying on manual input from authors upon submission to acquire metadata. However, making this manual task for  all past publications can be time-consuming. In fact, it might be impossible considering publications of all domains from past decades. As a result, developing an approach that allows automating this process is highly needed. Researchers have been constantly trying to solve this by proposing many approaches, many of which have achieved reliable results. However, these approaches were designed to tackle English scientific papers. The issue lies with German scientific publications \cite{Mexpub}. In contrast to English papers which follow a limited number of templates, German papers have a variety of different unseen templates. The latter introduces a new level of challenge for metadata extraction to deal with irregular layouts.

Research efforts that tackled the issue of metadata extraction have always relied on the natural language processing (NLP) methods to solve it, by either using rule-based approaches \cite{romary2015grobid} or by introducing machine learning \cite{alzaidy2019bi,an2017citation}. Even though they were able to achieve considerable results, the issue persisted with irregular layouts. In 2021, a research work \cite{Mexpub} proposed a novel method named MexPub. While common works used NLP to solve this issue, MexPub \cite{Mexpub} treated it as an image-vision problem. Instead of focusing on contextual features, they concentrate on visual features. This allowed for more flexibility when dealing with irregular layouts which are more common with German publications. Despite having improvements, MexPub lacks the ability to take contextual features into considerations, which can help in recognizing different classes such as emails, journals, and dates. 

Previously, combining multiple types of inputs has proven to be conducive to improve results for other fields. In this paper, we are proposing a multimodal deep learning approach that can employ advantages from both worlds of NLP and image-vision. The primary goal of this work is to emphasize on both contextual and visual features of publications in an effort to extract metadata reliably from German publications and increase the overall accuracy compared to previous state-of-the-art approaches. The ultimate aim of this approach is to extract segments of text which belong to one of the following classes: title, abstract, author, journal, address, email, date, DOI, and affiliation. 

In the following section, we outline the related works that addressed similar problems and the solutions they proposed. We introduce our proposed approach in detail in section 3, followed by a presentation of the experimental setup as well as a discussion of our obtained evaluation results in section 4.  Finally, we conclude the paper with the conclusion and future work in section 5.

\section{Related work}

\subsection{NLP}
According to \citet{Simes2009InformationET}, numerous previous old studies on metadata extraction relied on text and layout rules, addressing the issue using context-based classifiers such as Hidden Markov Models (HMMs) and similar approaches based on HMMs like Maximum Entropy Markov Models (MEMMs). However, these methods have several issues. First, the evaluation results on English papers were not reliable as many classes cannot be detected using only the features HMMs relies on. Second, HMMs struggle when dealing with multiple non-independent features because the states depend only on their immediate predecessor. Finally, HMMs and MEMMs are conducive to a weakness known as the label bias problem \cite{chaturvedi2018conditional}. One way of dealing with this issue is to use Conditional Random Fields (CRF) as proposed by GROBID \cite{romary2015grobid}, which is a machine learning library for extracting text from documents into a structured format such as XML. The approach relies on CRF in order to extract bibliographical header metadata (e.g., title, abstract, author, etc.). It is able to perform significantly well with the English language documents but when introduced to newly unseen templates, it faces challenges recognizing and extracting the metadata. 

Usually, with the previous approaches, the document goes through a sequential pipeline, mostly a two-phase process. Firstly, text segmentation and secondly the classification into one of the desired classes. The problem is that this type of approach is more perceptive to accumulated errors within the pipeline. To overcome such an issue \citet{8791225} proposed an end-to-end approach for reference segmentation and extraction from PDF documents. It learns the different characteristics of references to be used in a coherent scheme that reduces the error accumulation using a probabilistic approach.

\subsection{Computer Vision}

Computer vision is rarely used in document analysis. However,  recent experiments have shown that it is possible to obtain promising results with novel approaches in this area. For instance, \citet {Stahl:2018} proposed an approach titled DeepPDF that treats PDF as an image to label its texts to either "paragraph" or "non-paragraph" at a pixel-wise level. It uses the U-Net and a CNN architecture for semantic segmentation to classify each segment as “paragraphs“ or “non-paragraphs”.

Only a few approaches in the area of metadata extraction using deep learning have been proposed. For example, two of which had impressive results, SLLD from \citet{2010.11727} and MexPub from \citet{Mexpub}, proposed approaches which rely on the R-CNN \cite{1311.2524} architecture for object detection. SLLD implements Faster R-CNN \cite{7485869} with VoVNetV2-39 as a backbone (feature extraction and region selection) to detect the main region of the layout (e.g., title, authors, abstract of the text) while MexPub used Mask R-CNN \cite{8237584} with ResNeXt \cite{8100117} as a backbone with Feature Pyramid Networks (FPN). One main difference between SLLD and MexPub is the dataset. While MexPub's dataset is focusing on German literature, SLLD focuses more on English language documents. Both approaches have shown promising results for layout detection and metadata extraction from scientific literature. However, they still have some flaws. For example, SLLD fails to detect non-rectangular regions while MexPub has difficulties in precisely detecting small-sized patterns and detects multiple patterns mistakenly in some cases which lead to false positives. 

\begin{figure*}[ht]
  \includegraphics[width=\textwidth]{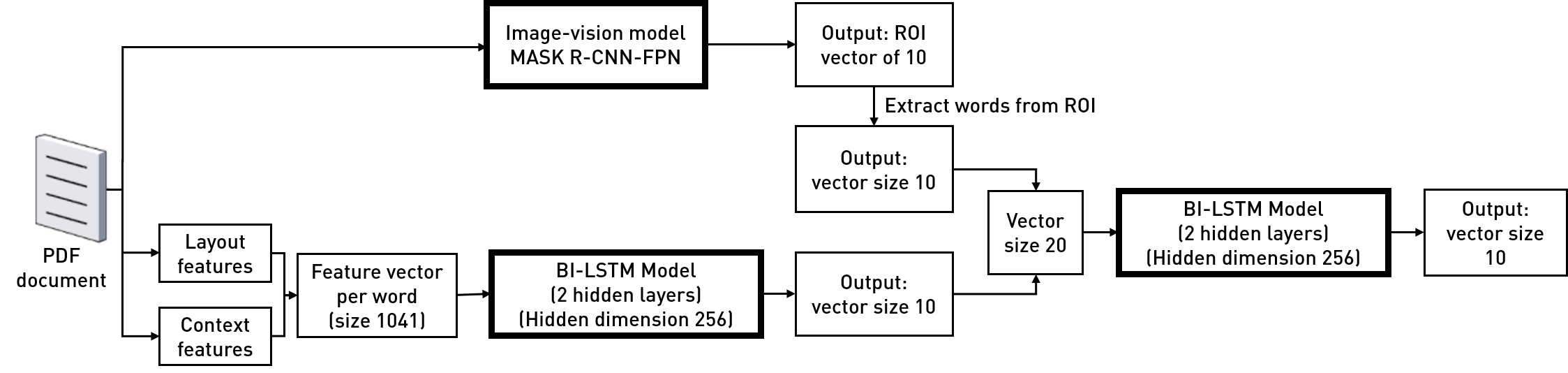}
  \caption{
      \textbf{Model Architecture} \\ This figure illustrates the multimodal architecture for  metadata extraction}
  \label{fig:architecture}
\end{figure*}

\subsection{Multimodality}

Multimodal deep learning has been used previously in many domains including audiovisual and image classification as it has shown it can improve the overall performance by combining multiple types of inputs and making them work together. Particular applications of multimodality in the field of metadata extraction have proven better results in comparison to unimodal counterparts as indicated by the works of \citet{Balasubramanian:2015} and \citet{Liu:2018}.

In the multimodal metadata extraction system for video lectures presented by \citet{Balasubramanian:2015}, both audio and video modalities are considered. This approach effectively utilizes the correlations between the audio transcripts and the content in the slides embedded in the video streams, to extract the defined metadata. A hybrid approach combining a Naive Bayes classifier and a rule-based refiner is used for the effective retrieval of metadata in a lecture video. By applying this multimodal methodology, in comparison to the audio-based approach, an improvement of 114.2\% is achieved when comparing precision, recall, and F-score for several lecture videos.

\citet{Liu:2018} was one of the firsts to introduce deep learning technology to the problem of metadata extraction. The multimodal approach for this extraction method can process both image data and text data as sources of information and does not need to design any classification feature.

Employing deep learning networks to model the image and text information of document headers respectively allowed to perform metadata extraction with little information loss. This approach presented two types of networks including Convolutional Neural Networks (CNNs) and Recurrent Neural Networks (RNNs) to handle both types of data. In the last stage, they combined those networks into a following Long-Short Term Memory (LSTM) to get the classifications of each line in paper headers. The overall system performance of this combined image and text model showed a great improvement after their integration: The images carry features such as location, font, and layout information, while the texts carry semantic and character arrangement information, and the model integration reduces the information loss in the metadata extraction process in general.

\section{Our approach}
We can decompose our overall architecture into three submodels: an NLP model, a computer vision model, and a classifier, as illustrated in Figure \ref{fig:architecture}. The following subsections describe the preprocessing steps and the detailed architecture of each submodel.

\subsection{Preprocessing}
Preprocessing is a crucial phase in which the acquired data is transformed to be the input for the model. For each of the documents, we extract the text of the first page using CERMINE \cite{CERMINE}. This tool was chosen because it is one of the most reliable ways to extract texts even for more challenging layouts and it enabled us to gather information about the geometric structure such as text position and font style in an easy manner. We only consider the first page of a document since the metadata information we are interested in typically exists within that scope.

After extracting the words of a document, each word runs through a layout and context feature extraction process. The textual layout feature vector consists of a set of 16 visual-spatial dependent features such as the horizontal or vertical space, the relative font size, the relative line number, the count of occurrences of specific special characters and multiple flags describing whether the text is in italic, bold or follows a specific common format such as date or email. 
These layout features on their own are not enough to represent the relationship between words and their classes. For instance, author names have a distinct meaning and usage compared to words that belong to an abstract. Consequently, we decided that there is a need to encode the context and the meaning of the words as well. To achieve this, we use the ELMO model by \citet{Che:2018} which encodes them into feature vectors of size 1025 to create a representation for each word. This ELMO model is based on the work of \citet{Peters:2018} and \citet{Gardner:2017} and trained on the German language, therefore it is better suited for our case compared to other models that have been trained on other languages.

The output of both feature extraction processes is concatenated to form a single feature vector of size 1041. This serves as input for the natural language processing submodel.

In contrast to the natural language processing phase, the image-vision does not rely on text but images. Each of the documents' first pages will be converted into an image which will be the input for the image-vision submodel.

\subsection{NLP Submodel}
\begin{figure}
\centering
  \includegraphics[width=9.5cm]{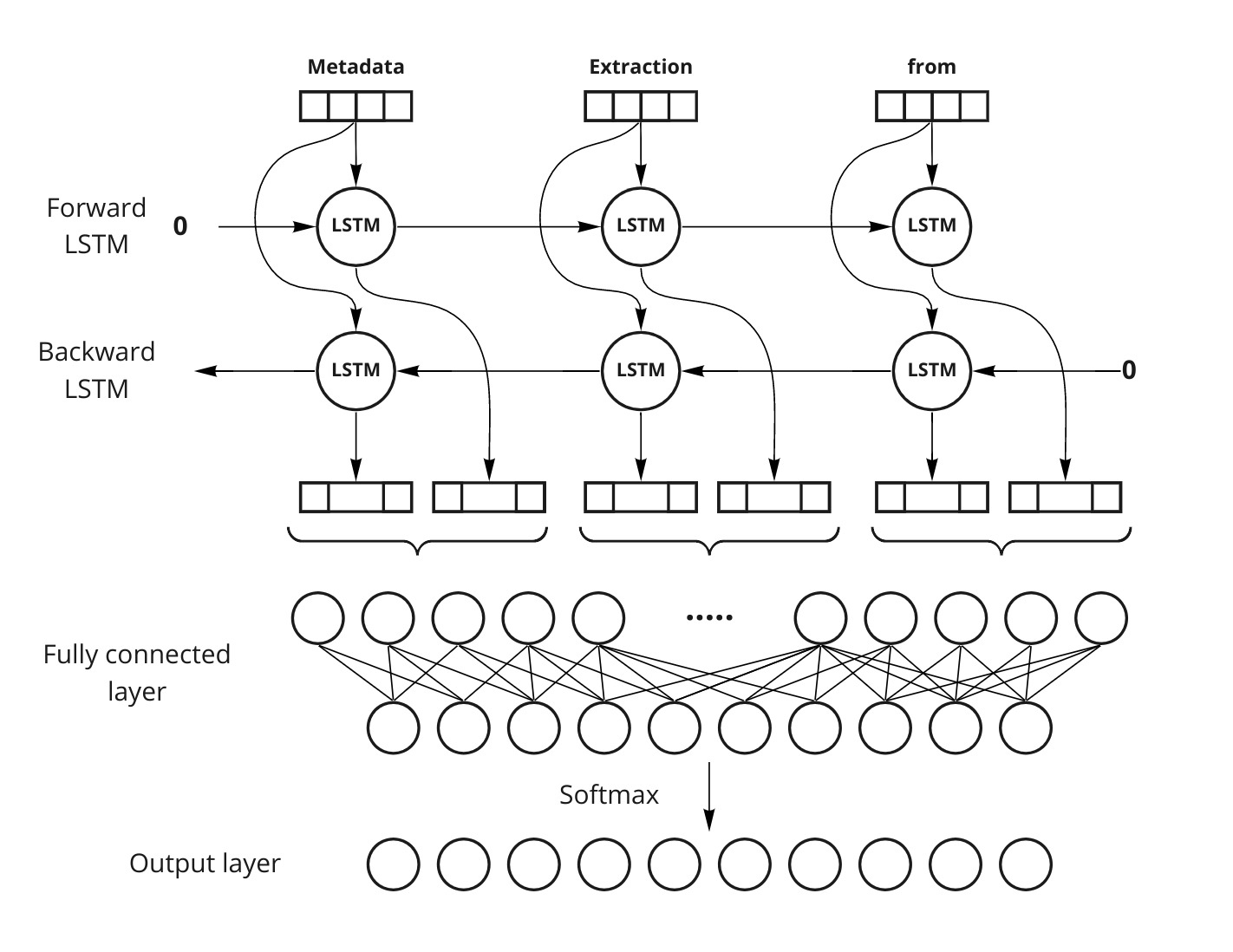}
  \caption{
      The NLP BiLSTM submodel architecture}
  \label{fig:nlp}
\end{figure}

According to \citet{8684825} BiDirectional Long-Short-Term Memory (biLSTM) is more efficient in learning the context within sentences than Long-Short-Term Memory (LSTM). Consequently, we have proceeded with the biLSTM as a model of choice for the NLP submodel. biLSTM essentially is two LSTM layers stacked together, as shown in Figure \ref{fig:nlp}. The first layer is a forward LSTM and the second is a backward LSTM. This allows the preservation of past and future information for a better contextual understanding.

The biLSTM model takes as input a word embedding vector of length 1041 at each time step. The embedding vector is composed of two parts, the first part of length 16 contains layout features (font size of the word, font style, spacing between the word and the line above/under, etc.) and the second part comprises the ELMO \cite{Che:2018} embedding results.

Our implementation of the biLSTM model comprises 2 hidden layers: forward and backward LSTM with 256 hidden dimensions each, then a fully connected layer of 512 input nodes, and 10 output nodes with a softmax activation function to obtain probability scores for the word belonging to each of the 10 classes (abstract, author, email, address, date, journal, affiliation, DOI, title, unclassified).

The output of the biLSTM is of length 512 (each LSTM layer's hidden dimension is 256) is fed to a classifier (fully connected neural network) with a softmax activation function that would result in an output of length 10, each output node $i$ represents the probability of a word being class $i$.

\subsection{Image vision Submodel}
In the image vision submodel we use the pre-trained model from MexPub \cite{Mexpub}, which is implemented with Mask R-CNN-FPN. Mask R-CNN is used for object instance segmentation and extended from Faster R-CNN. The improvements from its predecessor are a faster trainability as well as better overall results. By changing Region of Interest (RoI)-Pooling to RoIAlign, the model receives more precise region proposals for the objects to process. In addition, Mask R-CNN adds another two convolution layers that generate a mask for each RoI and segment at a pixel level.

The MexPub model is suited to our task since it aims to extract the same metadata and is already trained on German scientific papers. In addition, MexPub uses transfer learning by re-training the model from PubLayNet \cite{zhong2019publaynet}, which already has been trained on a large scientific literature dataset for classifying the following classes of title, text, list, table, and figure and is later fine-tuned to extract metadata classes instead.
\\

The submodel is built on Detectron2 \cite{wu2019detectron2}, a state-of-the-art detection and segmentation algorithm implemented in PyTorch. The library comes with pre-trained Mask R-CNN using ResNeXt-101 as a backbone with Feature Pyramid Network (FPN), and the process architecture is following Figure \ref{fig:imagevision}. Firstly, the PDF is converted into an image before feeding it into FPN to generate feature maps at a different scale. The output from FPN is fed into Region Proposal Network (RPN) to generate a bounding box that possibly contains the desired object. Then, a fully convolutional mask prediction branch is added to the head. Afterwards, ROI-Head obtains the output from FPN and RPN and use two fully convolution networks to generate bounding box and classification results. Finally, the output is filtered out by non-maximum suppression (NMS) to limit the results to the top hundred scoring detection boxes.
\\

In this submodel, we collect the bounding box and extract the text inside it, then aggregates the probability for all possible classes in that box from the model before feeding it into the classifier.

\begin{figure}
\centering
  \includegraphics[width=8cm]{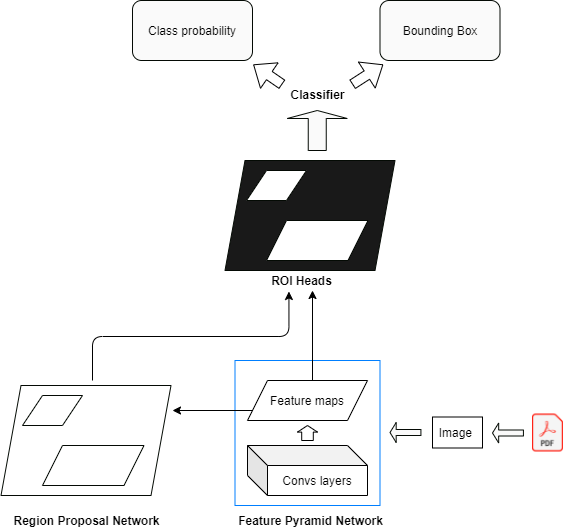}
  \caption{
      The image-vision submodel (MASK R-CNN FPN) architecture }
  \label{fig:imagevision}
\end{figure}

\subsection{Classifier}
\begin{figure}
\centering
  \includegraphics[width=9.5cm]{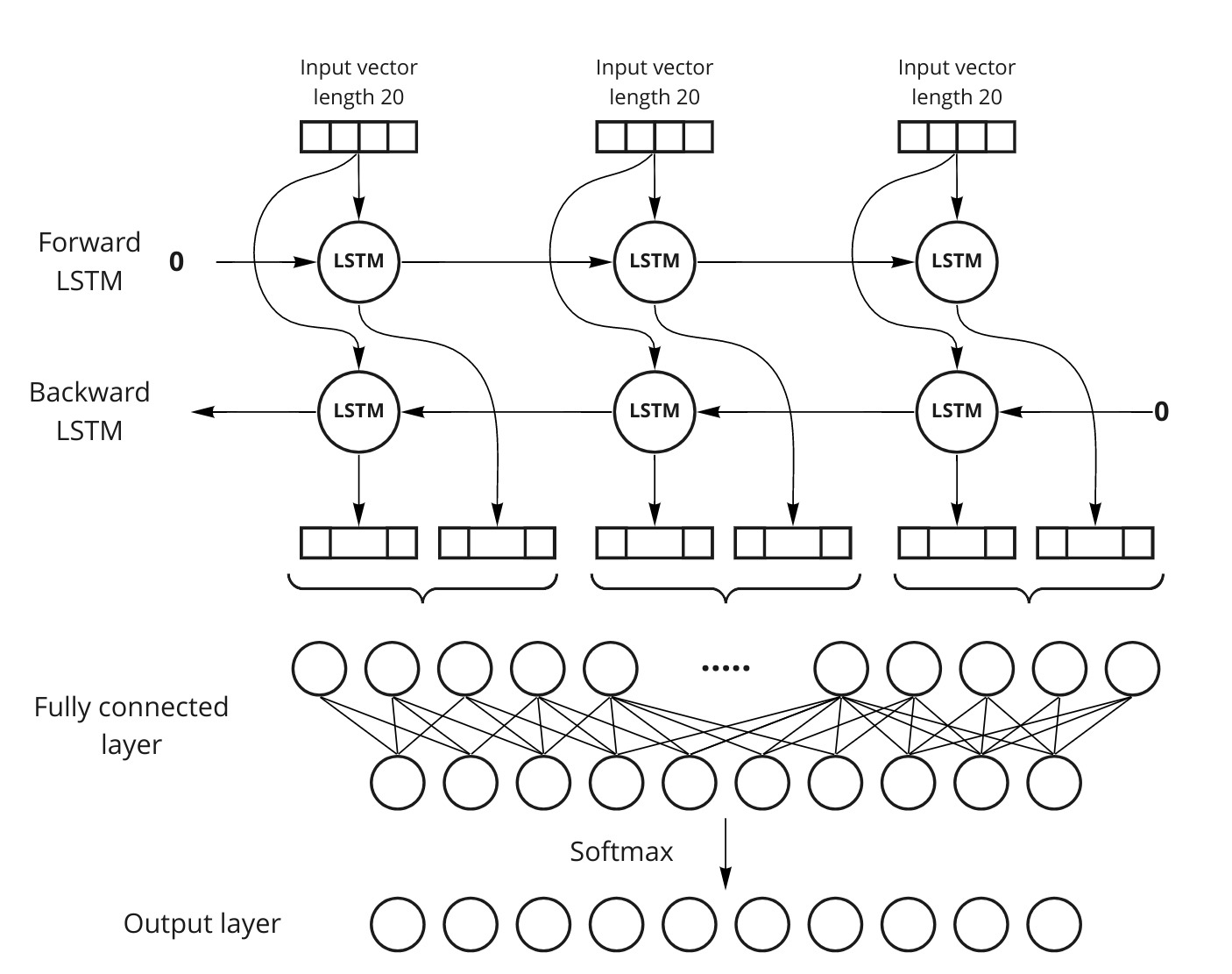}
  \caption{
      The Last BiLSTM submodel architecture}
  \label{fig:classifier}
\end{figure}

At the end of our architecture pipeline is our last submodel (classifier) that takes the output of the biLSTM and the processed output of Mask R-CNN model as input. The model of choice for this task is biLSTM because of the ability to learn patterns within sequences. The model takes a vector of length 20 (concatenation of both outputs of length 10 each) as input, and returns a probability distribution of length 10. The model consists of two stacked LSTM layers (forward and backward LSTMs) with 256 hidden dimensions as shown in Figure \ref{fig:classifier}. Finally, a fully connected layer containing 512 input nodes and 10 output nodes with a softmax activation function. The output gives a probability distribution of a word belonging to the 10 classes.

\section{Experiment}
\subsection{Data}
Due to the nature of our study that is focused on extracting metadata from German scientific papers only, acquiring labeled and suitable data is a challenging task. As a computer-vision approach does not necessarily rely on the language used, we assumed that a natural language processing model will be affected by the language in the training process. Therefore, it was important for us to filter documents that mainly used the German language. However, crawling a lot of German-only documents with all required metadata was difficult. Some of the metadata was often missing and many repositories of scientific publications do not offer an option to filter out by language.

In this paper, we used a collection of 300 manually annotated and 8518 generated documents. The dataset includes a variety of different non-standard layouts that is required to prevent our model on overfitting by learning exact specifics of the layouts and their text order.

The 300 documents are randomly selected from the SSOAR\footnote{https://www.gesis.org/ssoar/home} repository consisting of German scientific publications from the field of Social Science. We extracted and preprocessed the text of the first page for each document and labeled it on a word-by-word level based on our desired metadata classes. In case a word does not belong to any of these nine classes, we label them as unclassified (i.e. in the case of words of the first chapter such as the introductory section). The annotations for each document are saved in a CSV format.

However, manually annotating documents on a word-by-word level is complex and time-consuming. Therefore, we created a data generation process to enlarge our dataset. Due to this we randomly selected and extracted metadata records of German publications from SSOAR\footnotemark[1], media/rep/\footnote{https://www.mediarep.org/} and DBLP\footnote{https://dblp.org/xml/release/}. We used 13 commonly used layouts that we identified in our manually annotated dataset. Those layouts, among others, have been identified and used by MexPub \cite{Mexpub} as well. We generated 670 synthetic documents for each layout template by inserting the collected metadata at their respective positions in the template. Two of those layouts included an English abstract alongside the German abstract for which we did not have enough data crawled. Thus, we only generated 516 synthetic documents for each of those.

For each generated document we considered all components of the metadata as a whole, using all the extracted metadata from a particular record to generate a single document containing the same data.
This approach was chosen in order to preserve the context of the words in a document such that e.g. the meaning of the title and abstract relate to each other. We assumed that the context of words might be important for the model to learn. However, we did not try to fully randomize each field to generate different combinations of metadata, which would have enabled us to generate much more documents with the same amount of data, to test whether this affects the model’s performance. An example of a generated document using this process is shown in Figure \ref{fig:out}.

\subsection{Training}

\begin{figure}
\centering
  \includegraphics[width=8cm]{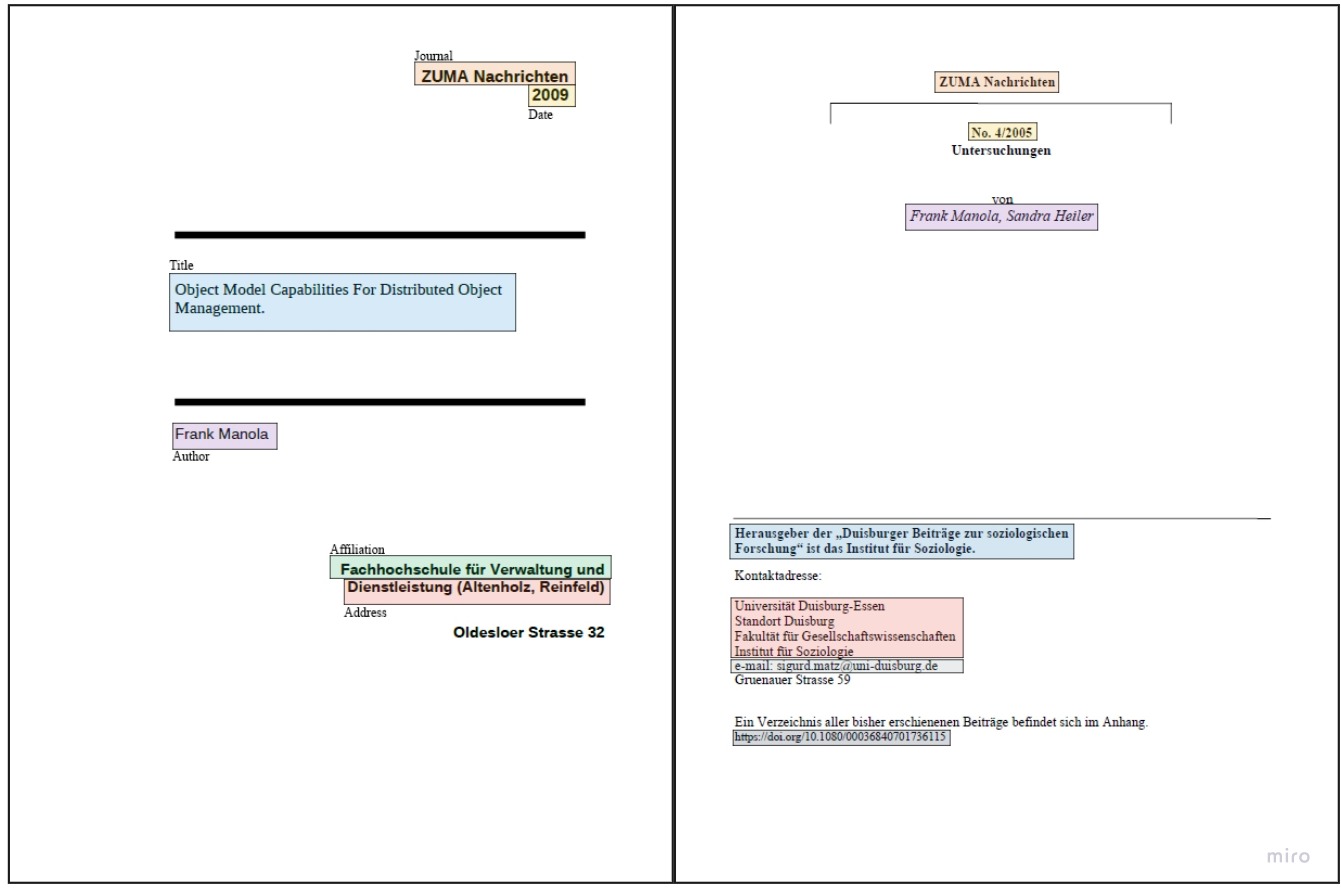}
  \caption{
      Example outputs of two randomly selected documents}
  \label{fig:out}
\end{figure}

To train and evaluate our model, we randomly split our training data into 70\% training, 15\% validation, and 15\% testing. Because of the high complexity of the architecture proposed, each of the submodels is trained independently using the processed data except for the image-vision model. The image-vision submodel \cite{Mexpub} is pre-trained on a similar dataset. Having the submodels trained separately enables us to identify and evaluate the issues of the individual submodels, making debugging easier as well as achieving a faster process overall. Figure \ref{fig:out} demonstrates the final model's output of two randomly selected documents from the testing set. This shows the approach's capability to enhance the performance of metadata extraction by using both contextual and visual features. Tables \ref{tab:nlp_eval}, \ref{tab:image_vis}. \ref{tab:final_eval} summarize the results obtained by this process for all submodels. We trained each of the biLSTM models with 300 iterations and a batch size of 2000. 

\begin{center}
    
\end{center}

\begin{table}[h!]
  \caption{Performance of the NLP submodel using the testing set}
  \label{tab:nlp_eval}
  \begin{tabular}{ccll}
    \toprule
     & Precision & Recall & F1-score\\
    \midrule
    \textbf{Overall}     & 0.835 & 0.819 & 0.827        \\
    \hline
    Abstract    & 0.934 & 0.922 & 0.928 \\
    Author      & 0.860 & 0.960 & 0.907 \\
    Email       & 0.938 & 0.949 & 0.943 \\
    Address     & 0.631 & 0.857 & 0.727 \\ 
    Date        & 0.980 & 0.800 & 0.881 \\
    Journal     & 0.900 & 0.737 & 0.810 \\
    Affiliation  &0.820 & 0.727 & 0.771 \\
    DOI         & 0.503 & 0.500 & 0.501 \\
    Title       &0.956 & 0.919 & 0.937 \\
  \bottomrule
\end{tabular}
\end{table}

\begin{table}[h!]
  \caption{Performance of the image-vision submodel using the testing set}
  \label{tab:image_vis}
  \begin{tabular}{ccll}
    \toprule
     & Precision & Recall & F1-score\\
    \midrule
    \textbf{Overall}     & 0.908 & 0.898 & 0.904 \\ 
    \hline
    Abstract    &0.961 & 0.913 & 0.936 \\
    Author      &0.932 & 0.940 & 0.936 \\
    Email       &0.918 & 0.891 & 0.904 \\
    Address     & 0.897 & 0.906 & 0.901 \\ 
    Date        &  0.934 & 0.931 & 0.932 \\
    Journal     &  0.960 & 0.895 & 0.926 \\
    Affiliation &  0.691 & 0.742 & 0.716 \\
    DOI         &   0.931 & 0.952 & 0.941 \\
    Title       &  0.950 & 0.920 & 0.935 \\ 
  \bottomrule
\end{tabular}
\end{table}

\begin{table}[h!]
  \caption{Performance of final model using the testing set}
  \label{tab:final_eval}
  \begin{tabular}{ccll}
    \toprule
     & Precision & Recall & F1-score\\
    \midrule
    \textbf{Overall}     & 0.944          & 0.902          & 0.923          \\ \hline
Abstract         & \textbf{0.989} & 0.962          & \textbf{0.975} \\
Author           & 0.962          & \textbf{0.984} & \textbf{0.973} \\
Email            & \textbf{0.985} & 0.953          & \textbf{0.969} \\
Address          & 0.922          & 0.940          & 0.931\\
Date             & 0.968          & \textbf{0.972} & \textbf{0.970} \\ 
Journal          & 0.971          & 0.937          & 0.954          \\
Affiliation      & \textbf{0.750} & \textbf{0.429} & \textbf{0.546} \\
DOI              & \textbf{0.986} & 0.967          & \textbf{0.976} \\
Title            & 0.963          & 0.961          & 0.962   \\
  \bottomrule
\end{tabular}
\end{table}

The proposed approach is able to achieve an overall F1-score of 92.18\% in the testing set when training all the components independently with 300 iterations. Achieving a high overall F1-score validates the model's proficiency in utilizing contextual and visual features to accurately extract metadata from German scientific publications. In fact, it was able to surpass the performance of the components (NLP and Image-vision model). The results show that the approach was able to perform better on extracting abstracts, authors, dates, and DOIs. The high F1-score for the abstract class could be explained by abstracts usually containing patterns that are unique across all layouts and typically containing 150-300 words that results in more space allocation within the paper, since both models are better at detecting larger patterns than smaller ones. Furthermore, in some cases we identified commonly used words such as  \textit{"journal"} that could act as marker words indicating a specific class. Similarly, the abstract is often preceded by words such as the German word \textit{"Zusammenfassung"}. As for the author class, using the contextual features helps to achieve this high performance because usually names are unique within the document which allows the model to better detect such names within a document page. For the email, DOI, and date classes the high F1-score is believed to be caused by  classes following a unique pattern across all papers, for example, the emails are usually within the format of "name@example.de". This unique pattern allows the model to learn much better in extracting emails; the same logic applies for dates and DOIs. 

However, along with the great evaluation results the approach has had, there are some weaknesses revealed by these results. The affiliation class appears to have a very low F1-score compared to other classes. We hypothesize that this low result is caused by the low occurrences and variety of the affiliation within the used dataset, therefore, adding more training data is advised in order to increase the performance of the model with such a class.

\subsection{Comparison with other approaches}
To have a better understanding of the performance of our model, we compared the approach against other state-of-the-art approaches GROBID \cite{romary2015grobid} and MexPub \cite{Mexpub}. To allow for a fair comparison, we randomly selected 300 documents from our testing set that our model has not seen during training. These documents will serve as input for all the approaches. For all methods, we extracted the metadata from the given 300 documents and calculated the cosine similarity between the extracted metadata and the ground truth metadata (labeled metadata) of the document. Note that the given 300 documents are labeled beforehand. If the cosine similarity was higher than 0.85, the extracted metadata is matching the correct one. The reason for allowing a dissimilarity of 0.15 is because there are cases where the extracted metadata misses a part of the text, but this does not necessarily mean the extraction is false. After extracting the metadata and calculating the cosine similarity for checking the F1-score is calculated to evaluate the approaches. Note that we have not included the evaluation for the classes \textit{date} and \textit{DOI} because some of the implemented approaches are not designed to extract them. Table \ref{tab:comparison} shows the obtained comparison results.

\begin{table}[h!]
  \caption{F1-score comparison between our approach, MexPub and GROBID}
  \label{tab:comparison}
  \begin{tabular}{cccc}
    \toprule
     & Our approach & MexPub & GROBID\\
    \midrule
    \textbf{Overall}     &\textbf{0.846} & 0.823          & 0.618 \\
    \hline
    Abstract    &    \textbf{0.923} & 0.910          & 0.821 \\
    Author      &  0.807          & \textbf{0.824} & 0.770 \\
    Email       &  0.844          & \textbf{0.901} & 0.624 \\
    Address     &  \textbf{0.870} & 0.821          & 0.324 \\
    Journal        &   \textbf{0.835} & 0.828          & 0.741 \\
    Affiliation     &  \textbf{0.679} & 0.535          & 0.240 \\
    Title &   \textbf{0.964} & 0.942          & 0.812 \\
  \bottomrule
\end{tabular}
\end{table}

Even though the proposed approach has slightly outperformed MexPub and GROBID in most classes, there are still two classes in which the approach did not perform better than MexPub. We hypothesize that the low results in the author and email classes compared to the ones from MexPub were due to the influence by the two submodels. If only one of the models outputs a bad result, then the last model is going to be influenced badly by this. As a result, we presume that implementing a way to add a confidence level to each of the submodels outputs would enhance the performance of the approach. By doing so, the last model will be able to weight each submodel output depending on its confidence level to prevent the influence of wrong results to the overall performance. Moreover, decreasing the model complexity could also help to solve the issue of error aggregation.

\section{Conclusion}
In this paper, we introduce an approach that automatically extracts metadata from German scientific papers using a deep learning multimodal approach. It is able to take advantage of layout and contextual features from a document by using both image-vision and natural language processing (NLP) models. Documents will be input into two different submodels (biLSTM, and pre-trained Mask R-CNN), the output of each of these models will be combined together into a single feature vector and fed into a final biLSTM model to classify words within a document into the appropriate metadata class (e.g., title, author, email, etc.). The overall performance of the approach was able to surpass other works by averaging an F1-score of 0.923 for all the classes. These results validate the assumption that including contextual and visual features increase the overall performance of metadata extraction for German scientific papers. 

In future work, it may be worth exploring more possibilities to decrease the complexity of the architecture to avoid the problem of error aggregation and allow for more generalization. Adding confidence level to each of the submodels could reduce the problem of error aggregation in the final submodel. In addition, using an additional conditional random field (CRF) layer at the end of the biLSTM classifier submodel could boost classification results as shown by \cite{Prasad:2018}. Moreover, due to the complexity of the architecture, increasing the training data with a larger amount of different layouts can further increase the performance especially with classes that got low results in the current approach. Finally, by collecting a larger dataset we could make a more in depth comparison between a pure NLP approach and a multimodal approach to better understand the strength and weaknesses of each one.

\section*{Venue}
This paper has been accepted and presented at the non-archival workshop: BiblioDAP@KDD2021\footnote{https://bibliodap.uni-koblenz.de/} \cite{boukhers2021bibliodap}


\bibliographystyle{ACM-Reference-Format}
\bibliography{bibliography}

\end{document}